\documentclass[11pt]{article}

\usepackage{graphicx,amssymb,amsbsy}
\usepackage{natbib}

\newcommand\Real{\mbox{Re}} 
\newcommand\Imag{\mbox{Im}} 
\newcommand\Rey{\mbox{\textit{Re}}}  

\newsavebox{\astrutbox}
\sbox{\astrutbox}{\rule[-5pt]{0pt}{20pt}}

\def\pr{{\partial}}
\def\eps{{\epsilon}}

\begin{document}

\centerline{\Large{\bf Instability of diverging and converging flows in an annulus}}

\vskip 5mm
\centerline{\bf Konstantin Ilin\footnote{Department of Mathematics, University of York,
Heslington, York YO10 5DD, UK. Email address for correspondence: konstantin.ilin@york.ac.uk} and 
Andrey Morgulis\footnote{Department of Mathematics, Mechanics and Computer Science, The Southern Federal University, Rostov-on-Don, and South Mathematical Institute, Vladikavkaz Center of RAS, Vladikavkaz,
Russian Federation}}

\begin{abstract}
\noindent
The stability of two-dimensional diverging and converging flows in an annulus between two permeable cylinders is examined. The basic flow
is irrotational and has both the radial and azimuthal components. It is shown that for a wide range of the parameters of the problem,
the basic flow is unstable to small two-dimensional perturbations. The instability is inviscid and oscillatory and persists
if the viscosity of the fluid is taken into consideration.
\end{abstract}

\section{Introduction}
In this paper we present a simple example of an instability of a steady inviscid flow in an annulus between
two permeable circular cylinders. Despite the utter simplicity of the basic flow, this instability
appears to be overlooked by other researchers.
The basic flow is irrotational and rotationally symmetric. Both the radial and the azimuthal
components of the velocity are inversely proportional to the radius.
We show that if the ratio of the azimuthal component of the velocity to its
 radial component is larger than
a certain critical value, then the flow is unstable to small two-dimensional perturbations. The instability is oscillatory: the neutral modes
represent azimuthal travelling waves. The most striking feature of the instability is that it is present not only in the diverging flow
but also in the converging flow. This contradicts the common view that diverging flows are always unstable while converging
flows are stable, which originates in the studies of the Jeffery-Hamel flow \citep[e.g.,][]{Goldshtik1991,Shtern,Drazin1998}. 

The stability of viscous flows between permeable rotating cylinders with a radial flow to three-dimensional perturbations
had been studied by many authors
\citep[e.g.,][]{Bahl, Chang, Min, Kolyshkin, Kolesov}.  One of the main aims of these studies was to determine the effect of the radial flow
on the stability
of the circular Couette-Taylor flow, and the general conclusion was that the radial flow changes the stability properties of the flow:
both a converging radial flow and a sufficiently strong diverging flow have a stabilizing effect on the Taylor vortex instability,
but when a divergent flow is weak, it has a destabilizing effect \citep{Min, Kolyshkin}. Since the above papers had been focused on the
effect of the radial flow on the known instability, it remained unclear whether
the radial flow itself can induce instability for flows which are stable without it. The results described below may help to answer this question.
It should be mentioned that the purely azimuthal flow with the velocity inversely proportional to the radius is stable not only to two-dimensional perturbations \citep[see][]{Drazin} but also to three-dimensional perturbations (this can be deduced from the sufficient condition for stability given
by \citet{Howard}). The results of our study show that this stable flow becomes unstable if the radial flow is present and the weaker this radial
flow is, the more unstable the flow becomes.

The outline of the paper is as follows. In Section 2, we formulate the problem. Section 3 contains the inviscid stability analysis
of the diverging flow.
In Section 4 we show that the inviscid stability results of Section 3 represent a valid asymptotic of the corresponding viscous problem
in the limit of high Reynolds numbers. In Section 5, we briefly discuss the stability of the converging flow.
Finally, conclusions are presented in
Section 6.

\section{Formulation of the problem}\label{sec:problem}
We consider two-dimensional inviscid incompressible flows in an annulus $D$ between two concentric circles
with radii $r_{1}$ and $r_{2}$ ($r_2 > r_1$). The circles are permeable for the fluid and there is a constant volume flux $2\pi Q$ of
the fluid through the annulus (the fluid is pumped into the annulus at the inner circle and taken out at the outer circle).
Suppose that
$r_1$ is taken as a length scale, $r^2_{1}/Q$ as a time scale, $Q/r_{1}$ as a scale for the velocity and $\rho Q^2/r_{1}^2$ for the pressure
where $\rho$ is the fluid density. Then the two-dimensional Euler equations, written in non-dimensional variables, have the form
\begin{eqnarray}
&&u_{t}+ u u_{r} + \frac{v}{r}u_{\theta} -\frac{v^2}{r}= -p_{r} ,  \label{1} \\
&&v_{t}+ u v_{r} + \frac{v}{r}v_{\theta} +\frac{u v}{r}= -\frac{1}{r} \, p_{\theta} ,  \label{2} \\
&&\frac{1}{r}\left(r u\right)_{r} +\frac{1}{r} \, v_{\theta}=0.  \label{3}
\end{eqnarray}
Here $(r,\theta)$ are the polar coordinates, $u$ and $v$ are the radial and azimuthal components of the velocity and $p$ is the pressure.
It is known that if there is a non-zero flow of the fluid through the boundary, it is necessary to prescribe additional boundary conditions
on the part of the boundary where the fluid enters the flow domain. What conditions should be added is a subtle question and there are
several answers that lead to mathematically correct initial boundary value problems \citep[see, e.g.,][]{Monakh, MorgYud}. We
will use the boundary condition for the tangent component of the velocity, which at the first approximation corresponds to
the condition at a porous cylinder \citep[see][]{Joseph} and for which the corresponding mathematical problem is well-posed \citep[e.g.,][]{Monakh}.
So, our boundary conditions are
\begin{equation}
u\!\bigm\vert_{r=1}=1, \quad u\!\bigm\vert_{r=a}=1/a,
\quad v\!\bigm\vert_{r=1}=\gamma,  \label{4}
\end{equation}
where $a=r_2/r_1$ and $\gamma$ is a constant that represents the ratio of the azimuthal velocity to the radial velocity at the inner circle.

Problem (\ref{1})--(\ref{4}) has the following simple rotationally-symmetric solution:
\begin{equation}
u(r,\theta)=1/r, \quad v(r,\theta)=\gamma/r.  \label{5}
\end{equation}
In the next section we investigate the stability of this steady flow.

\section{Inviscid stability analysis}\label{sec:inviscid_stability}

We consider a small perturbation
$(\tilde{u}, \tilde{v}, \tilde{p})$ in the form of the normal mode
\begin{equation}
\{\tilde{u}, \tilde{v}, \tilde{p}\} = Re\left[\{\hat{u}(r), \hat{v}(r), \hat{p}(r)\} e^{\sigma t + in\theta}\right]  \label{3.1}
\end{equation}
where $n\in\mathbb{Z}$. This leads to the eigenvalue problem:
\begin{eqnarray}
&&\left(\sigma +  \frac{in\gamma}{r^2} + \frac{1}{r} \, \pr_{r} \right) \hat{u}
-\frac{1}{r^2} \, \hat{u} +\frac{2\gamma}{r^2} \, \hat{v} = -\hat{p}_{r} ,  \label{3.2} \\
&&\left(\sigma +  \frac{in\gamma}{r^2} + \frac{1}{r} \, \pr_{r} \right) \hat{v}
+\frac{1}{r^2} \, \hat{v}  = -\frac{in}{r} \, \hat{p} ,  \label{3.3} \\
&&\frac{1}{r}\left(r \hat{u}\right)_{r} +\frac{in}{r} \, \hat{v}=0 \label{3.4}
\end{eqnarray}
and
\begin{equation}
\hat{u}(1)=0, \quad \hat{u}(a)=0,
\quad \hat{v}(1)=0.  \label{3.5}
\end{equation}
First we note that the eigenvalue problem (\ref{3.2})--(\ref{3.5}) has no nontrivial solution for $n=0$. Indeed,
Eq. (\ref{3.4}) for $n=0$ and the boundary conditions for $\hat{u}$ imply that $\hat{u}=0$.
Equation (\ref{3.3}) yields $\hat{v} = C \, r^{-1} \, e^{-\sigma r^2/2}$
where $C$ is an arbitrary constant. Substitution of this into the boundary condition
$\hat{v}(1)=0$ leads to the conclusion that $C=0$.

Now consider eigenvalue problem (\ref{3.2})--(\ref{3.5}) for $n\neq 0$. It is
convenient to introduce the stream function $\hat{\psi}(r)$ such that
\[
\hat{u}=\frac{in}{r} \, \hat{\psi}(r), \quad \hat{v}=-\hat{\psi}'(r).
\]
After elimination of the pressure, we obtain
\begin{eqnarray}
&&\left(\sigma +  \frac{in\gamma}{r^2} + \frac{1}{r} \, \pr_{r} \right)L \hat{\psi}=0 ,  \label{3.6} \\
&&\hat{\psi}(1) = \hat{\psi}(a)=0, \quad \psi'(1)=0,  \label{3.7}
\end{eqnarray}
where
\begin{equation}
L \hat{\psi}= \hat{\psi}''+\frac{1}{r}\hat{\psi}'-\frac{n^2}{r^2}\hat{\psi} . \label{3.8}
\end{equation}
It follows from (\ref{3.6}) that
\begin{equation}
L \hat{\psi}= C e^{-h(r)} \label{3.9}
\end{equation}
where $C$ is a constant and $h(r)=\sigma r^2/2+i n \gamma \ln r$.
The general solution of (\ref{3.9}) is
\begin{equation}
\hat{\psi} = \frac{C_{1}}{r^n}+C_{2} r^n + \frac{C}{2n}\int\limits_{1}^{r}
\left(r^n s^{-n+1}- r^{-n}s^{n+1}\right)e^{-h(s)} \,  ds \label{3.10}
\end{equation}
where $C_{1}$ and $C_{2}$ are arbitrary constants.
Substitution of (\ref{3.10}) into the boundary conditions (\ref{3.7}) yields a system of linear equations for
$C_{1}$, $C_{2}$ and $C$. The requirement that this system has a non-trivial solution results in the dispersion relation
for $\sigma$:
\begin{equation}
D(\sigma, n, a, \gamma)\equiv \int\limits_{1}^{a}
\left(a^n r^{-n+1}- a^{-n}r^{n+1}\right)e^{-\sigma r^2/2-i n \gamma \ln r} \, dr=0 . \label{3.11}
\end{equation}

\begin{figure}
\begin{center}
\includegraphics*[height=6cm]{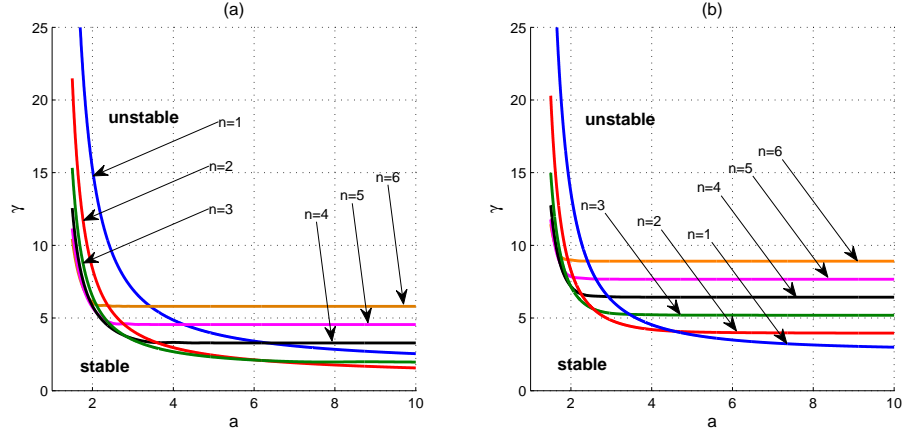}
\end{center}
\caption{Neutral curves for $n=1,\dots,6$. The region above each curve is where the corresponding mode is unstable.
(a) - diverging flow, (b) - converging flow.}
\label{stab_bound}
\end{figure}

A few conclusions can be made just by looking at the dispersion relation.
First, for the purely radial flow ($\gamma=0$), it shows that (i) there are no real eigenvalues (because the integrand in (\ref{3.11}) is non-negative for all
$r\in[1,a]$) and (ii) the eigenvalues appear in complex conjugate pairs (i.e. if $\sigma$ is an eigenvalue, so is its complex conjugate $\bar{\sigma}$).
Second, it is easy to see that
\begin{eqnarray}
&&\overline{D(\sigma, n, a, \gamma)}=-D(\bar{\sigma}, -n, a, \gamma), \label{3.12} \\
&&D(\sigma, n, a, \gamma)=-D(\sigma, -n, a, -\gamma). \label{3.13}
\end{eqnarray}
These relations imply 
it suffices to consider
only positive $n$ and $\gamma$.

\begin{figure}
\begin{center}
\includegraphics*[height=5.5cm]{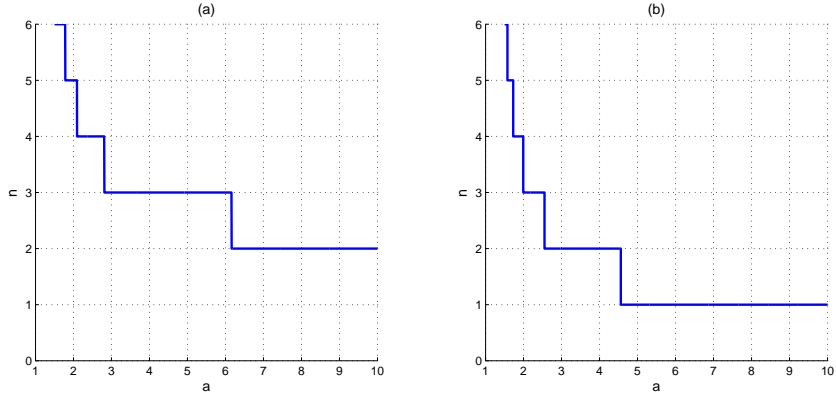}
\end{center}
\caption{Azimuthal wave number $n$ of the most unstable mode as a function of $a$. (a) - diverging flow, (b) - converging flow.}
\label{most_unstable_mode}
\end{figure}

Numerical evaluation of (\ref{3.11}) shows that
(i) the purely radial flow ($\gamma=0$) is stable for any $a>1$, i.e. $\Real(\sigma)< 0$ for all azimuthal modes, and
(ii) for each azimuthal mode, when $\gamma$ increases, the egenvalues in the upper half plane ($\Imag(\sigma)>0$) move to the right and
the egenvalues in the lower half plane ($\Imag(\sigma)<0$) move to the left, and
there is a critical value $\gamma_{c}>0$ of parameter $\gamma$ at which one of the eigenvalues crosses the imaginary axis, so that
\[
\Real (\sigma) > 0 \ \ {\rm for} \ \ \gamma >\gamma_{c} \  \ {\rm and} \ \ \Real (\sigma) < 0 \ \
{\rm for} \ \ \gamma <\gamma_{c}.
\]
Neutral curves ($\Real (\sigma) = 0$) on the $(a,\gamma)$ plane  are shown in Fig. \ref{stab_bound}(a) for $n=1,\dots,6$.
For each azimuthal mode, the critical circulation $\gamma_{c}$ is a decreasing function of $a$ and tends to a limit value as $a\to\infty$.
The calculations also show that, for each $n$, $\gamma_{c}\to\infty$ as $a\to 1$. For large $a$, the first mode that
becomes unstable when $\gamma$ increases from $0$ (we will call it the most unstable mode) has
the azimuthal wave number $n=2$. When $a$ decreases, first the mode with $n=3$ becomes the most unstable one, then the mode with
$n=4$, and so on.
The wave number $n$ of the most unstable mode versus $a$ is shown in Fig. \ref{most_unstable_mode}(a). 
The jumps in $n$ correspond to the intersection points of the neutral curves in Fig. \ref{stab_bound}.

\begin{figure}
\begin{center}
\includegraphics*[height=6cm]{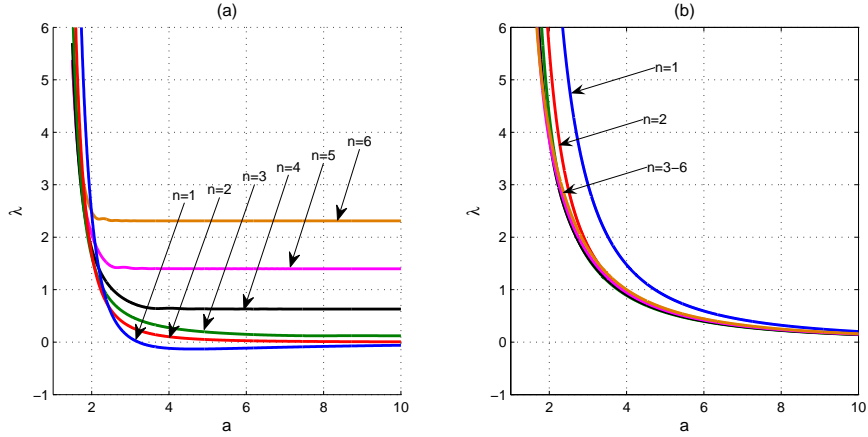}
\end{center}
\caption{The phase speed $\lambda=-Im(\sigma)/n$ of the neutral modes versus $a$ for $n=1,\dots,6$.
(a) - diverging flow, (b) - converging flow.}
\label{phase_speed}
\end{figure}

This instability is oscillatory. The neutral modes represent azimuthal waves travelling with the phase speed $\lambda=-\Imag(\sigma)/n$
in the counterclockwise direction.
The graphs of $\lambda$ versus $a$ for $n=1,\dots,6$ are shown in Fig. \ref{phase_speed}(a). For $n=2,\dots,6$, the phase speed is positive
for all $a$, while for the neutral mode with $n=1$, it changes sign: it is positive for relatively small $a$ and negative (though small
in magnitude) for large $a$.

\section{Effect of viscosity}\label{sec:viscous_stability}
The steady flow (\ref{5}) is also the solution of the Navier-Stokes equations that satisfies (\ref{4}) and the additional
boundary condition
\begin{equation}
v\!\bigm\vert_{r=a}=\gamma/a. \label{4.1}
\end{equation}
This solution corresponds to the situation when the inner and outer cylinders rotate anticlockwise with angular
velocities $\gamma$ and $\gamma/a^2$ respectively.
The aim of this section is to show that for sufficiently high Reynolds numbers the unstable inviscid modes found in section 3
give a good approximation to the corresponding viscous modes.

The viscous counterpart of Eq. (\ref{3.6}) is
\begin{equation}
\left(\sigma +  \frac{in\gamma}{r^2} + \frac{1}{r} \, \pr_{r} \right)L \hat{\psi}= \frac{1}{\Rey} \, L^2 \hat{\psi} , \label{4.2}
\end{equation}
where $\Rey =Q/\nu$ is the Reynolds number
($\nu$ is the kinematic viscosity of the fluid).
Equation (\ref{4.2}) and the boundary conditions (cf. (\ref{3.7}))
\begin{equation}
\hat{\psi}(1) = 0, \quad \hat{\psi}'(1)=0, \quad \hat{\psi}(a)=0, \quad \hat{\psi}'(a)=0 \label{4.3}
\end{equation}
represent an eigenvalue problem for $\sigma$. In what follows we will construct an asymptotic expansion
of the solution to this eigenvalue problem in the limit $\Rey\to\infty$. It is known \citep[e.g.,][]{Temam, Yudovich2001, Ilin2008} that
for high Reynolds numbers, a problem like this involves a boundary layer at the part of the boundary
where the fluid leaves the domain. So, we assume that the asymptotic expansion has the form
\begin{eqnarray}
&& \sigma=\sigma_{0}+\Rey^{-1}\sigma_{1}+\Rey^{-2}\sigma_{1}+\dots ,  \label{4.4} \\
&& \hat{\psi}=\hat{\psi}_{0}(r)+\Rey^{-1} [\hat{\psi}_{1}(r)+\phi_{0}(\eta)]+\Rey^{-2} [\hat{\psi}_{2}(r)+\phi_{1}(\eta)]+\dots  \label{4.5}
\end{eqnarray}
Here $\eta=(a-r)\Rey$ in the boundary layer variable. Functions $\hat{\psi}_{k}(r)$ ($k=0,1,\dots$) represent
the regular part of the expansion, and $\phi_{k}(\eta)$  ($k=0,1,\dots$) give us boundary layer corrections
to the regular part. The latter ones are needed to satisfy the additional boundary condition that appears in
the viscous problem. We assume that the boundary layer part rapidly decays outside thing boundary layer near
$r=a$:
\begin{equation}
\phi_{k}(\eta) =o(\eta^{-s}) \quad {\rm as} \quad \eta\to\infty ,    \label{4.6}
\end{equation}
for every $s>0$ and for each $k=0,1,\dots$

The regular part of the expansion is obtained by substituting (\ref{4.4}) and the formula
\begin{equation}
\hat{\psi}=\hat{\psi}_{0}(r)+\eps \hat{\psi}_{1}(r)+\eps^2 \hat{\psi}_{2}(r)+\dots  \label{4.7}
\end{equation}
into Eq. (\ref{4.2}). This yields
\begin{eqnarray}
&&\left(\sigma_{0} +  \frac{in\gamma}{r^2} + \frac{1}{r} \, \pr_{r} \right)L \hat{\psi}_{0}= 0, \label{4.8} \\
&&\left(\sigma_{0} +  \frac{in\gamma}{r^2} + \frac{1}{r} \, \pr_{r} \right)L \hat{\psi}_{1}= -\sigma_{1}L \hat{\psi}_{0} +
 L L \hat{\psi}_{0}, \quad {\rm etc.} \label{4.9}
\end{eqnarray}
To obtain boundary conditions at $r=1$, we
substitute (\ref{4.7}) into the first two boundary conditions
(\ref{4.3}). This yields
\begin{equation}
\hat{\psi}_{k}(1) = \hat{\psi}'_{k}(1)=0  \label{4.10}
\end{equation}
for each $k=0,1,\dots$ For the boundary conditions at $r=a$, the boundary layer part must be taken into account. So we substitute
(\ref{4.5}) into the last two conditions (\ref{4.3}) and use the relation $\pr_{y}=-\Rey \, \pr_{\eta}$. As a result, we get
\begin{eqnarray}
&&\hat{\psi}_{0}(a)=0, \quad \hat{\psi}'_{0}(a)-\phi'_{0}(0)=0, \label{4.11} \\
&&\hat{\psi}_{1}(a)+\phi_{0}(0)=0, \quad \hat{\psi}'_{1}(a)-\phi'_{1}(0)=0, \quad {\rm etc.} \label{4.12}
\end{eqnarray}
Comparing Eqs. (\ref{4.8}), (\ref{4.10}) (for $k=0$)  and the first of the conditions (\ref{4.11}) with
(\ref{3.6}) and (\ref{3.7}), we conclude that, at the leading order, we have the inviscid eigenvalue problem.
Note that at this stage we cannot satisfy the second of the conditions (\ref{4.11}),  and that is why we need a boundary layer at $r=a$.

To derive equations for the boundary layer part of the expansion, we substitute (\ref{4.5})
into Eq. (\ref{4.2}) and take into account that $\hat{\psi}_{k}$ ($k=0,1,\dots$)
satisfy (\ref{4.8}), (\ref{4.9}). Then we make the change of variables
$r=a-\Rey^{-1} \, \eta$, expand every function of $a-\Rey^{-1}  \, \eta$ in Taylor's series at $\Rey^{-1}=0$
and, finally, collect terms of the equal powers in $\eps$. In the leading order, we obtain
\begin{equation}
- \frac{1}{a} \, \pr_{\eta}^3 {\phi}_{0}=  \pr_{\eta}^4 {\phi}_{0}.  \label{4.13}
\end{equation}
This should be solved subject to the condition of decay at infinity and the second condition
(\ref{4.11}) that can be written as
\begin{equation}
\phi'_{0}(0)=\hat{\psi}'_{0}(a). \label{4.14}
\end{equation}
The solution of
(\ref{4.13}), (\ref{4.14}) is
\begin{equation}
\phi_{0}(\eta)=-a \hat{\psi}'_{0}(a)e^{-\eta/a}. \label{4.15}
\end{equation}
Note that the boundary layer does not affect the leading order eigenvalue $\sigma_{0}$. It can be shown that the boundary layer will
affect the first order viscous correction $\sigma_{1}$, but we will not compute it here.

\section{Converging flow}\label{sec:converging}
In the previous two sections we have found that the diverging flow (\ref{5}) can be unstable in the framework of the inviscid theory
and that the instability persists if the viscosity is taken into consideration. The natural question to
ask is whether this instability
occurs only in diverging flows or a similar converging flow is also unstable?
To answer this question we consider the converging flow
\begin{equation}
u(r,\theta)=-1/r, \quad v(r,\theta)=\gamma/r,  \label{5.1}
\end{equation}
which differs from (\ref{5}) only by the sign of the radial component of the velocity.
This is a rotationally-symmetric solution of Eqs. (\ref{1})--(\ref{3}) that satisfies the boundary conditions
\begin{equation}
u\!\bigm\vert_{r=1}=1, \quad u\!\bigm\vert_{r=a}=\frac{1}{a},
\quad v\!\bigm\vert_{r=a}=\gamma .  \label{5.2}
\end{equation}
Note that now
the third boundary condition (for the azimuthal component of the velocity) is imposed
at $r=a$ (i.e. on the part of the boundary where the fluid flows into the domain).

A repetition of the analysis of Section 3 leads to the following dispersion relation for the eigenvalue $\sigma$:
\begin{equation}
D_1(\sigma, n, a, \gamma)\equiv \int\limits_{1}^{a}
\left(r^{-n+1}- r^{n+1}\right)e^{\sigma r^2/2 + i n \gamma \ln r} \, dr=0 . \label{5.3}
\end{equation}
This dispersion relation is very similar to the one we had before (cf. (\ref{3.11})) and satisfies the same symmetry relations
(\ref{3.12}) and (\ref{3.13}). 
The neutral curves in the $a \, $-$\gamma$ plane are shown in Fig. \ref{stab_bound}(b) for the modes
with $n=1,\dots,6$.
The wave number $n$ of the most unstable mode as a function of $a$ is shown in Fig. \ref{most_unstable_mode}(b).
Unlike the diverging flow, here the mode with $n=1$ is the most unstable one for sufficiently large $a$. The phase speed $\lambda=-\Imag(\sigma)/n$
is shown in Fig. \ref{phase_speed}(b). In contrast with the diverging flow, the phase speed for the converging flow
is positive for all $n\geq 1$ and its magnitude
is almost the same for all modes with $n\geq 3$.

\section{Conclusions}

We have shown that a simple inviscid irrotational flow between two permeable cylinders are unstable to small two-dimensional perturbations.
The instability is oscillatory and persist if the viscosity of the fluid is taken into account. It is a genuinely new instability in the sense that
there is no instability if either the radial flow or the azimuthal flow is absent.

As we discussed in Section 4, the inviscid
flow (\ref{5}) is also an exact solution of the Navier-Stokes equations corresponding
to a particular case of the flow between permeable rotating cylinders when
$\Omega_1/\Omega_2=r_{2}^2/r_1^2$. In the general steady rotationally symmetric viscous flow, the radial component of the velocity
is the same as before, while the azimuthal component is $v=A \, r^{1+\Rey} + B \, r^{-1}$ where $A$ and $B$ are constants depending on
$\gamma_1=\Omega_1 r_1^2/Q$, $\gamma_2=\Omega_2 r_2^2/Q$, $a=r_{2}/r_1$ and \Rey. In the limit
of high Reynolds numbers this flow is well approximated by (\ref{5}) everywhere except for a boundary layer at the flow outlet. 
This fact suggests that the instability described above may be relevant to the whole family of rotationally symmetric viscous flows. 
This, nowever, requires a
further analysis and is a topic of a continuing investigation.

There are many interesting question that are left unanswered in the present paper. For instance,
it is known that in problems with nonzero flow through the boundary,
the stability may depend on the boundary conditions imposed at the inlet and the outlet (see, e.g., \cite{Chomaz}).
The question that arises in this context: will the same flow be unstable if different boundary conditions are imposed?
Another interesting question: what are the stability properties of the flow (\ref{5}) with respect to three-dimensional perturbations?
At the moment, these are open problems for future investigations.

We are grateful to Professor V. A. Vladimirov for helpful discussions.

\bibliographystyle{jfm}


\end{document}